 \documentclass[twocolumn,showpacs,preprintnumbers,pre]{revtex4}

\def\ADD#1{{\textcolor{blue}{#1}}} 
\def\ADD2#1{{\textcolor{red}{#1}}} 

\def\DEL#1{{\textcolor{green}{}}} 
\usepackage{amssymb,amsmath,amsfonts}
\usepackage{graphicx,graphics}
\DeclareGraphicsExtensions{.jpg,.jpeg,.pdf,.png,.mps,.eps,.ps,.gif}
\usepackage{color}
\usepackage{framed}
\graphicspath{{matlab/}}

\usepackage{xr}
\usepackage{epstopdf}

\newcommand{\rem}[1]{}
\DeclareMathAlphabet{\mathbi}{OML}{cmm}{b}{it} 

\newcommand{\non}{\nonumber}

\newcommand{\bx}{\mathbi{x}}

\newcommand{\bel}{\begin{equation}\label}
\newcommand{\ee}{\end{equation}}
\newcommand{\beq}{\begin{eqnarray}\label} 
\newcommand{\eeq}{\end{eqnarray}} 
\newcommand{\bc}{\begin{center}} 
\newcommand{\ec}{\end{center}} 
\newcommand{\ben}{\begin{enumerate}}
\newcommand{\een}{\end{enumerate}}
\newcommand{\bit}{\begin{itemize}}
\newcommand{\eit}{\end{itemize}}
\newcommand{\I}{\int_{\mathcal{V}}}
\newcommand{\bB}{\mbox{\boldmath$B$}}
\newcommand{\bdA}{\mbox{\boldmath$A$}}
\newcommand{\bdD}{\mbox{\boldmath$D$}}
\newcommand{\bdQ}{\mbox{\boldmath$Q$}}

\newcommand{\bdb}{\mbox{\boldmath$b$}}

\newcommand{\bhr}{\mbox{\boldmath$\hat{r}$}}

\newcommand{\br}{\mathbi{r}}

\newcommand{\bdf}{\mathbi{f}}

\newcommand{\bu}{\mbox{\boldmath$u$}}
\newcommand{\bv}{\mathbi{v}}

\newcommand{\bom}{\mbox{\boldmath$\omega$}}
\newcommand{\bphi}{\mbox{\boldmath$\phi$}}

\newcommand{\bj}{\mbox{\boldmath$j$}}

\newcommand\shalf{\ensuremath{{\scriptstyle\frac{1}{2}}}}

\newcommand\squart{\ensuremath{{\scriptstyle\frac{1}{4}}}}

\newcommand{\Dmpm}{\left[D_{m}^{\pm}\right]}
\newcommand{\Dmmp}{\left[D_{m}^{\mp}\right]}

\newcommand{\Dopm}{\left[D_{1}^{\pm}\right]}
\newcommand{\Domp}{\left[D_{1}^{\mp}\right]}
\newcommand{\bz}{\mathbi{z}}

\begin{document}
\title{Depletion of Nonlinearity in Magnetohydrodynamic Turbulence: Insights
from Analysis and Simulations\footnote{Postprint version of the manuscript published in Phys. Rev. E 93, 043104 (2016).}} 
\author{J. D. Gibbon$^1$, A. Gupta$^2$, G. Krstulovic$^3$, R. Pandit$^4$, 
H. Politano$^5$, Y. Ponty$^3$, A. Pouquet$^{6,7}$, G. Sahoo$^2$, and J. Stawarz$^7$} 
\affiliation{
$^1$Department of Mathematics, Imperial College London, London SW7 2AZ, UK.\\
$^{2}$Department of Physics, University of Rome `Tor Vergata', 00133 Roma, Italy.\\
$^{3}$Laboratoire Lagrange, Universit\'e C\^ote d'Azur, Observatoire de la C\^ote d'Azur, 
CNRS, Blvd de l'Observatoire, CS 34229, 06304 Nice cedex 4, France.\\
$^{4}$Centre for Condensed Matter Theory, Indian Institute of Science, Bangalore, 560 012, India. \\
$^{5}$Laboratoire Dieudonn\'e, Universit\'e de Nice  Sophia-Antipolis, France.  \\
$^{6}$National Center for Atmospheric Research, P.O. Box 3000, Boulder, CO 80307, USA.\\
$^{7}$Laboratory for Atmospheric and Space Physics, University of Colorado, Boulder, CO 80303, USA.
 }
\begin{abstract}
It is shown how suitably scaled, order-$m$ moments, $D_m^{\pm}$, of the Els\"asser  vorticity fields in three-dimensional
magnetohydrodynamics (MHD) can be used to identify three possible regimes for solutions of the MHD equations 
with magnetic Prandtl number $P_M = 1$. These vorticity fields are defined by $\bom^{\pm} = \mbox{curl}\,\bz^{\pm} = 
\bom\pm \bj$,  where $\bz^{\pm}$ are Els\"asser variables, and where $\bom$ and $\bj$ are, respectively, the fluid vorticity 
and current density. This study follows recent developments in the study of three-dimensional Navier-Stokes fluid turbulence 
[Gibbon \textit{et al.} Nonlinearity 27, 2605 (2014)].  Our mathematical results are then compared with those from a variety 
of direct numerical simulations, which demonstrate that all solutions that have been investigated remain in \textit{only 
one of these regimes} which has depleted nonlinearity. The exponents $q^{\pm}$ that characterize the inertial range 
power-law dependencies of the $\bz^{\pm}$ energy spectra, $\mathcal{E}^{\pm}(k)$, are then examined, and bounds are 
obtained. Comments are also made on\,: (a) the generalization of our results to the case $P_M \neq1$ and (b) the relation 
between $D_m^{\pm}$ and the order-$m$ moments of gradients of magnetohydrodynamic fields, which are used to 
characterize intermittency in turbulent flows.
\end{abstract}
\maketitle

\section{Introduction}\label{sec:intro}

Intermittency is widespread in nature\,: its characterization is a central problem in turbulence~\cite{frisch_95,
kerr_85,meneveau_91,SreenivasanAntonia97,boffetta_08,arneodo_08,donzis_08,ishihara_09,pandit_09}, nonequilibrium statistical
mechanics, the production and storage of wind and solar energy, and the behaviors of market crashes, and of several critical 
phenomena~\cite{bramwell_98,sornette_06}. It has also been studied extensively in fluid 
turbulence~\cite{frisch_95,meneveau_91,SreenivasanAntonia97,boffetta_08,arneodo_08,donzis_08,ishihara_09,pandit_09} and in magnetohydrodynamic (MHD) turbulence~\cite{biskamp_00,mueller_00,mininni_07,mininni_09,sahoo_11,poletto2013sun}, often
by using order-$p$ structure functions of fields such as the velocity and, in MHD, the magnetic field. An example is the (longitudinal) 
velocity ($\bu$) structure function, 
\beq{sfn1}
S_{p}(r) &\equiv& \langle [\delta u(r)]^p\rangle\,,\non\\
\delta u(r) &\equiv& [\bu(\bx + \br) - \bu(\bx)] \cdot \bhr\,,
\eeq
which scales as
\bel{sfn2}
S_{p}(r) \sim r^{\zeta_p}
\ee
for $ \eta_d \ll r \ll L$, where $\eta_d$ is the dissipation length scale below which 
viscous dissipation is significant, $L$ is the large length scale at which energy is injected into the fluid, and the multiscaling exponents 
$\zeta_p$, which are nonlinear, monotone increasing functions of $p$, characterize the intermittency~\cite{frisch_95}. Simple
scaling is obtained if $\zeta_p$ depends linearly on $p$, as in the phenomenological approach (K41) of Kolmogorov~\cite{K41}. 

To determine $\zeta_p$ is a challenging task~\cite{boffetta_08,arneodo_08,pandit_09}, which is especially difficult for time-dependent 
structure functions~\cite{mitra_04,ray_08} or MHD turbulence~\cite{biskamp_00,mueller_00,mininni_07,mininni_09,sahoo_11,poletto2013sun}. Therefore, we explore other signatures 
of intermittency. For three-dimensional (3D) fluid turbulence Refs.~\cite{JDGCMS11,donzis_13,gibbon_14} have introduced a new way 
of analyzing direct numerical simulations (DNSs) to obtain fresh insights into suitably scaled (see below), order-$m$ moments $D_m$ of 
the vorticity $\bom = \nabla \times \bu$. These studies show the following: (a) on theoretical grounds, three regimes, I, II, and III, are 
possible, with the $D_m$ ordered in different ways (Fig. 1, Ref.~\cite{gibbon_14}); but (b) \textit{only regime I is observed in a wide 
variety of DNSs}~\cite{donzis_13,gibbon_14}. Regime I has sufficiently depleted nonlinearity so that a global attractor exists, 
provided the solutions remain in this region, as they do in all the DNSs examined so far from this point of view.

The analog of the above theoretical framework is developed for the case of 3D MHD turbulence. Then the behaviors of $D_m^{\pm}$ 
are examined -- the 3D MHD counterparts of $D_m$ in \cite{donzis_13,gibbon_14} --  in a variety of DNSs, which have been carried out independently by different groups, to obtain new insights into the depletion of nonlinearity here. It is found that 3D MHD turbulence 
is like its fluid-turbulence counterpart inasmuch as all solutions remain in \textit{only one regime, with depleted nonlinearity, in a 
large variety of DNSs}.  The implications of our results are also examined for the exponents $q^{\pm}$ that characterize the power-law 
inertial range dependencies of the energy spectra $\mathcal{E}^{\pm}(k)$ of the Els\"asser variables on the wave number $k$. 

The remainder of this paper is organized as follows. In \S\ref{num} the MHD equations are introduced and our numerical methods 
are summarized. \S\ref{Math} contains the mathematical analysis of these equations. \S\ref{sec:spectra} is devoted to the energy 
spectra that emerge from these calculations. \S\ref{Con} contains the principal conclusions of the paper. The technical details of 
some of our calculations have been relegated to Appendices A, B and C. 

\section{Model and numerical methods}\label{num}

\subsection{The equations in Els\"asser variables}

The velocity $\bu$ and magnetic field $\bdb$ can be combined into the Els\"asser variables 
\bel{Elsass1}
\bz^\pm = \bu \pm \bdb\,.
\ee
Then the incompressible 3D MHD equations are
\beq{zfielda} 
(\partial_{t} + \bz^{\mp} \cdot \nabla)\bz^{\pm} 
&=&   \nu_{+}\nabla^{2}\bz^{\pm} +  \nu_{-}\nabla^{2} \bz^{\mp}\non\\
&-&\nabla {\cal P} + \bdf^{\pm}\,,        
\eeq
where $\nabla \cdot \bz^{\pm} =0$, ${\cal P}$ is the total pressure, $\nu_\pm = \shalf(\nu\pm\eta)$,  and $\nu$ 
and $\eta$ are, respectively, the kinematic viscosity and the magnetic diffusivity, whose ratio yields the magnetic 
Prandtl number $P_M=\nu/\eta$.  The two forcing functions, $\bdf^{\pm}(\bx)$, are defined by
\bel{Elsass2}
\bdf^\pm(\bx) = \bdf_{u} \pm \bdf_{b}\,.
\ee
which are absent in decaying MHD turbulence. $\bj =\nabla\times \bdb$ is the current density. The mean magnetic 
field $\bB_{0}$ in zero in our simulations. The following notation will be used for spatial and temporal averages\,: 
\bel{not1a}
\left<\cdot\right>_{V} = L^{-3}\I\cdot\,dV\,,
\ee
\bel{not1b}
\left<\cdot\right>_{T} = T^{-1}\int_{0}^{T}\cdot\,dt\,,
\ee
with the $L^2$-spatial norm represented by
\bel{not2}
\|\cdot\|_{2} = \left(\I |\cdot|^{2}\,dV\right)^{1/2}\,.
\ee
The Taylor-microscale Reynolds number $R_{\Lambda}$ is defined as 
\bel{TMS1}
R_{\Lambda} = u_{rms}\nu^{-1} \left(\frac{\left<\bu^{2} + \bdb^{2}\right>_{V}}
{\left<\bom^{2} + \bj^{2}\right>_{V}}\right)^{1/2}\,,
\ee
with $u_{rms}$ the root-mean-square velocity. Our DNSs of the 3D MHD equations use a periodic cubic box and a pseudo-spectral 
method~\cite{biskamp_00,mueller_00,mininni_07,mininni_09,sahoo_11} with large-scale initial conditions, and in some cases, 
forcing (Table \ref{table1} ). All our numerical simulations are fully de-aliased.

For ideal 3D MHD (i.e., $\nu_\pm=0, \, \bdf^\pm=0$) the invariants are the  energies 
\bel{en1}
E_{\pm} = \shalf\left<\bz_{\pm}\cdot \bz_{\pm}\right>_{V} = E_{T} \pm H_C
\ee
together with the magnetic and cross-helicities
\bel{hel1}
H_{M}= \left< \bdA \cdot \bdb \right>_{V}\,,\qquad H_C = \left< \bu \cdot \bdb\right>_{V}\,,
\ee
where  the vector potential $\bdA$ is related to $\bdb$ by 
\bel{Adef}
\bdb = \nabla\times \bdA\,,
\ee
and the total energy is
\beq{hel2}
E_{T} &=& \shalf\left< \bu \cdot \bu  +  \bdb\cdot \bdb \right>_{V}\non\\
&=& E_{u} + E_{b}\,.
\eeq
The relative rates of magnetic and cross-helicity are also defined as 
\bel{hel3}
\sigma_{m}=\cos (\bdA, \bdb)\,,\qquad \sigma_{C}=\cos (\bu, \bdb)\,,
\ee 
with $|\sigma_{c,m}| \le 1$. These represent the degree to which the fields are aligned and 
they are also measures, global or point-wise, of the strength of nonlinearities in MHD.

By defining the two combinations of the vorticity and the current as
\bel{ompm}
\bom^\pm = \bom\pm \bj\,,
\ee
it is shown in Appendix \ref{app1} that $\bom^\pm$ evolve according to (with $P_{M}=1$)
\beq{omz}
(\partial_{t} &+& \bz^{\mp}\cdot\nabla) \bom^{\pm}  - \bom^{\mp}\cdot\nabla\bz^{\pm}   - \nu \Delta\bom^{\pm}\\
&=&  \bom^{\mp}\times\bom^{\pm} + \sum_{i=1}^{3}\partial_{i}\bz^{\pm}\times\partial_{i}\bz^{\mp} + \nabla \times \bdf^{\pm} 
\non\,.  
\eeq
The two terms on the right-hand side stem from the equation for the current; the labels $i = 1,\,2$, and $3$ refer, 
respectively, to $x,\,y$, and $z$. 

In the ideal case, the constraints that follow from conservation laws involve mixed $(\bu,\,\bdb)$ correlators~\cite{politano_03,basu_14}. 
In the absence of a strong uniform magnetic field $\bB_0$, magnetic fluctuations at a scale comparable to that of the system, 
$\bB_L$, play a role equivalent to that of $\bB_0$ for the small scales, provided there is sufficient scale separation, i.e., for 
high-Reynolds-number flows. It has been argued in \cite{mininni_07} that measurable anisotropy develops for scales smaller than 
the Taylor scale based on $\bB_L$. Therefore, the inertial-range energy spectrum can be of either Kolmogorov (K41) or 
Iroshnikov-Kraichnan (IK) forms, depending on the cross-correlation.  Dimensional analysis gives 
\bel{IK1}
\zeta^{IK}_p=p/4\,,
\ee
if the model of Iroshnikov and Kraichnan (IK)~\cite{iroshnikov_63, rhk_65} is used and $\sigma_C=0$\,; or 
\bel{K41a}
\zeta^{K41}_p=p/3\,,
\ee
if K41~\cite{K41,sahoo_11} is used. Appendix \ref{P:F} discusses some of these scaling arguments in a 
phenomenonogical manner. Moreover, 
\bel{Espec}
\mathcal{E}^{\pm}(k) \sim \left\{
\begin{array}{l}
 k^{-3/2}~~~(IK)\\
k^{-5/3}~~~(K41)\,.
\end{array}
\right.
\ee
Some models \cite{grauer_94, politano_95a} and DNS results~\cite{politano_98b,mueller_03,sahoo_11} 
indicate that the departure from linear scaling, be it of the IK or K41 forms, is stronger in 
3D MHD turbulence than in 3D Navier-Stokes (NS) turbulence, which suggests a 
depletion of nonlinearity by virtue of the tendency of alignment or anti-alignment of 
$\bu$ and $\bdb$~\cite{servidio_08a,align1}. 

\subsection{Description of runs}

Table \ref{table1} contains the parameters for the runs analyzed in this paper. All runs have been 
performed in three dimensions by using periodic boundary conditions, no imposed external magnetic 
field and a magnetic Prandtl number $P_M$ of unity, except for the pm-runs\,; no modeling of the 
small scales is employed. For the sd-runs (spin-down) the Reynolds number is varied. The initial 
condition for the spin-down runs sd is the three-dimensional Orszag-Tang vortex \cite{stawarz_12}, 
with added phase shifts to set $\sigma_C \simeq -0.21$ initially. The Aa-Ae runs are high-resolution 
forced runs \cite{bigot_08, homann_14}, with a constant velocity and magnetic forcing for which 
all the modes in the first two Fourier shells  are kept constant. From the Aa to Ae runs, the resolution 
increases with the Reynolds number. The tg-runs are forced in both the velocity and induction equations. 
In these runs, the four-fold symmetries of the Taylor-Green vortex extended to MHD are implemented. 
Moreover, the three runs  have different resulting energy spectra (IK, K41, and $k^{-2}$), although they 
have the same ideal invariants but with different cross-correlations \cite{krstulovic_14}. Finally, the pm 
runs have a fixed viscosity, but variable magnetic diffusivities and thus allow for extending the analysis 
to the case of $P_M\not= 1$ (see \cite{sahoo_11}). 
\begin{widetext}
\begin{table*}
 \caption{\label{table1}
Parameters for our direct numerical simulations. $k_{\rm max}=N/3$ is the maximum resolved wavenumber at 
grid resolution $N$ (the standard 2/3 de-aliasing rule). $\Lambda_{T}$ and $R_\Lambda$ are defined in (\ref{TMS1}). 
$\sigma_C$ and  $\sigma_M$ are the relative rates of cross-helicity and magnetic helicity, respectively.   $\lambda^\pm$ 
are the parameters extracted from the data for high $m$ (subscript max) and low $m$ (subscript min) 
(See Fig.\ref{Fig:AllRuns}, column 3).}
\begin{ruledtabular}
\begin{tabular}{ccccccccccccccccc}
Run&$N$ & $R_{\Lambda}$ & $\Lambda_T$& $P_M$   & $\sigma_C$& $\sigma_M$& $\lambda^+_{\rm min}$ & 
$\lambda^+_{\rm max}$ & $\lambda^-_{\rm min}$ & $\lambda^-_{\rm max}$ \\
\hline
sd1& 128  & $14$ & $0.27$ & $1$ & $-0.27$  & $-0.22$  & $1.096$ & $1.158$ & $1.101$ & $1.169$  \\
sd2& 256 & $21$ & $0.20$ & $1$  & $-0.27$   & $-0.23$ & $1.103$ & $1.165$ & $1.116$ & $1.186$ \\
sd3& 512 & $30$ & $0.15$ & $1$  &  $-0.27$  & $-0.24$ & $1.111$ & $1.171$ & $1.129$ & $1.197$ \\
sd4& 768 & $45$ & $0.11$ & $1$  &  $-0.26$  & $-0.24$ & $1.121$ & $1.184$ & $1.141$ & $1.206$\\
\hline
Aa& 512 & $35 $ & $0.098 $ & $1$     & 0.019 &  0.003  &   1.049 &   1.150 &  1.049 & 1.156 \\
Ab& 1024 & $ 54 $ & $0.074$ & $1$   & 0.017 &  0.004 &    1.057  &   1.197  &  1.060 & 1.195  \\
Ac& 2048 & $ 120$ & $0.036 $ & $1$  & 0.011 &   Data Not Available     &     1.076  &   1.167  &  1.076 &  1.176  \\
Ad& 2048& $ 161$ & $0.027$ & $1$   &  0.009 &   Data Not Available   &    1.074  &   1.168  &  1.073 & 1.157 \\
Ae& 4096 & $341 $ & $0.014$ & $1$   & 0.010 &   Data Not Available    &    1.070 &    1.163 & 1.072 &   1.174 \\
\hline
tgi& 1024 & $100$ & $0.066$ & $1$ & $0$& $0$	& $1.121$ &$1.196$ & $1.117$ &$1.197$\\
tga& 1024& $83$ & $0.084$ & $1$&$0$& $0$		& $1.161$ &$1.202$ & $1.138$ &$1.202$\\
tgc& 1024& $110$ & $0.056$ & $1$ &$\sim 0.05$& $0$	& $1.084$ &$1.183$ & $1.089$ &$1.175$\\
\hline
pm01& 512& $240$ & $0.14$  & $0.1$  & $0.122$ & $0.0047$ & $1.078$ & $1.238$ & $	1.078$ & $1.234$\\
pm02& 512& $140$ & $0.10$  & $1.0$  & $0.075$ & $0.0049$ & $1.070$ & $1.171$ & $	1.069$ & $1.160$\\
pm03& 512& $80$ & $0.06$  & $10$  & $0.226$ & $0.0077$ & $1.053$ & $1.149$ & $	1.052$ & $1.158$\\
\end{tabular} 
\end{ruledtabular} 
\end{table*}
\end{widetext}

\section{Mathematical analysis\label{Math}}

The generalization of the analysis of Refs.~\cite{donzis_13,gibbon_14} for the 3D NS equations is now described in the case of the 3D MHD equations. The relevant partial differential equations (PDEs) in Els\"asser variables are (\ref{zfielda}) and (\ref{omz}). Two spatially and 
temporally averaged velocities, $U^{\pm}$, based on $\bz^{\pm}$, are defined as
\bel{Upmdef}
U^{\pm2} = L^{-3}\left<\|\bz^{\pm}\|_{2}^{2}\right>_{T}\,.
\ee
In turn, the $U^{\pm}$ allow us to define two Reynolds numbers 
\bel{twoRe}
Re_{\pm}= LU^{\pm}/\nu\,.
\ee
The Reynolds numbers are based on average velocities, while two Grashof numbers $Gr_{\pm}$ are based on the 
forcing functions $\bdf^{\pm}(\bx)$:
\bel{Grdef1}
Gr_{\pm} = L^{3/2}\|\bdf^{\pm}\|_{2}/\nu^{2}\,.
\ee
For the class of forcing functions spectrally concentrated around a single length-scale ($\ell = L$ for the purposes 
of this paper), a relation exists between $Gr_{\pm}$ and $Re_{\pm}$ for solutions of (\ref{zfielda}) derived through the method of Doering-Foias \cite{doering_02} (see Appendix \ref{app2}) where it has been 
shown that 
\bel{GrRe}
Gr_{\pm} \leq c\,Re_{\pm}\left(Re_{\mp} + 1\right)\,,\qquad Gr_{\pm} \gg 1\,.
\ee
The main variables used in this paper are $L^{2m}$-norms of the vorticity field, defined in such a way that 
each has the dimension of a frequency\,:
\bel{Omdef1}
\Omega_{m}^{\pm}(t) = \left(L^{-3}\I |\bom^{\pm}|^{2m}dV\right)^{1/2m}\,.
\ee
By H\"older's inequality, the $\Omega_{m}^{\pm}$ are naturally ordered such that
\bel{Omorder}
\Omega_{1}^{\pm} \leq \Omega_{m}^{\pm} \leq \Omega_{m+1}^{\pm}\,.
\ee
If a signal has no intermittency, then the $\Omega_{m}^{\pm}$ will be packed close together, whereas a strongly 
intermittent signal will cause them to spread out widely. The following scaling was first introduced in work 
on the 3D Navier-Stokes equations \cite{JDGCMS11,donzis_13,gibbon_14} and will be followed here\,:
\bel{Dmdef1}
D_{m}^{\pm} = \left(\varpi_{0}^{-1}\Omega_{m}^{\pm}\right)^{\alpha_{m}}\,,
\ee
where the exponent $\alpha_{m}$ is defined as
\bel{alphadef}
\alpha_{m} = \frac{2m}{4m-3}\,,
\ee
and where $\varpi_{0} = \nu L^{-2}$ is the box frequency. The $\alpha_{m}$-scaling comes from symmetry 
considerations. The ordering of the $\Omega_{m}^{\pm}$ in (\ref{Omorder}) does not necessarily hold for the 
$D_{m}^{\pm}$ as $\alpha_{m}$ is decreasing with respect to $m$. The $D_{m}^{\pm}$ are the main variables 
to be used. Under the assumption that (\ref{zfielda}) has a solution we now look at the evolution of $D_{1}^{\pm}$\,:
\begin{widetext}
\beq{D1pmA}
\shalf \varpi_{0}^{-1}\dot{D}_{1}^{\pm} &\leq& - L^{-1}\varpi_{0}^{-2}\I |\nabla\bom^{\pm}|^{2}dV 
+ L^{-3}\varpi_{0}^{-3}\I \left| \bom^{\pm}\cdot\left(\bom^{\mp}\cdot\nabla\bz^{\pm}\right)\right|\,dV\nonumber\\ 
&+& L^{-3}\varpi_{0}^{-3}\sum_{i=1}^{3}\I \left| \bom^{\pm}\cdot\left(\partial_{i}\bz^{\pm}
\times\partial_{i}\bz^{\mp}\right)\right|\,dV + Gr_{\pm}D_{1}^{\pm1/2}\,.
\eeq
To estimate the first nonlinear term in (\ref{D1pmA}) we write ($ 1 < m < \infty$)
\beq{nlt1}
\I \left|\bom^{\pm}\cdot\left(\bom^{\mp}\cdot\nabla\bz^{\pm}\right)\right|\,dV &\leq& 
\I \left|\bom^{\pm}\right|^{\frac{2m-3}{2(m-1)}}\left|\bom^{\pm}\right|^{\frac{1}{2(m-1)}}
\left|\bom^{\mp}\right|^{\frac{2m-3}{2(m-1)}}\left|\bom^{\mp}\right|^{\frac{1}{2(m-1)}}
\left|\nabla\bz^{\pm}\right|\,dV\nonumber\\
&\leq& c_{1,m}\left(\I|\bom^{\pm}|^{2}\,dV\right)^{\frac{2m-3}{4(m-1)}}\left(\I|\bom^{\pm}|^{2m}\,dV\right)^{\frac{1}{4m(m-1)}}\\
&\times& \left(\I|\bom^{\mp}|^{2}\,dV\right)^{\frac{2m-3}{4(m-1)}}\left(\I|\bom^{\mp}|^{2m}\,dV\right)^{\frac{1}{4m(m-1)}}\nonumber
\left(\I|\bom^{\pm}|^{2m}\,dV\right)^{1/2m}\,.
\eeq
Note that the sum of the five exponents in the latter expression is unity. For the last term we have invoked the inequality\footnote{At $m=1$, 
we have equality with $c_{1}=1$. The case $m=\infty$ needs a logarithmic correction.}, which requires a Riesz transform in its proof, namely,
\bel{nltex1}
\|\nabla\bz^{\pm}\|_{2m} \leq c_{2,m}\|\bom^{\pm}\|_{2m}\,,
\ee
provided $1 \leq m < \infty$. Then (\ref{nlt1}) becomes
\beq{nlt2}
L^{-3}\varpi_{0}^{-3}
\I \left|\bom^{\pm}\cdot\left(\bom^{\mp}\cdot\nabla\bz^{\pm}\right)\right|\,dV &\leq &
c_{3,m} \Dopm^{\frac{2m-3}{4(m-1)}}\Dmpm^{\frac{1}{2\alpha_{m}(m-1)}}\nonumber\\
&\times& \Domp^{\frac{2m-3}{4(m-1)}}\Dmmp^{\frac{1}{2\alpha_{m}(m-1)}}
\Dmpm^{\frac{1}{\alpha_{m}}}\,.\nonumber\\
&\leq &
c_{3,m} \Dopm^{\frac{2m-3}{4(m-1)}}\Dmpm^{\frac{2m-1}{2\alpha_{m}(m-1)}}\nonumber\\
&\times& \Domp^{\frac{2m-3}{4(m-1)}}\Domp^{\frac{1}{2\alpha_{m}(m-1)}}.
\eeq
As in the 3D Navier-Stokes equations, this estimate of the nonlinearity is too strong for the dissipation terms. 
However, what was observed in computations of the 3D Navier-Stokes equations is that it displays numerically 
much weaker behavior than the estimate equivalent to (\ref{nlt2}) \cite{gibbon_14}. This can be 
measured by numerically tracking $D_{m}$ in terms of $D_{1}$, the equivalent of which for 3D MHD is\footnote{The 
simple H\"older inequality $\Omega_{1}^{\pm} \leq \Omega_{m}^{\pm}$ translates to $D_{1}^{\pm\alpha_{m}/2} 
\leq D_{m}^{\pm}$ which, in turn, implies that the lower bound $D_{1}^{\pm\alpha_{m}/2}$ is equivalent to 
$\lambda^{\pm} = 1$.}\,:
\bel{numdep1}
D_{m}^{\pm} \leq \left[D_{1}^{\pm}\right]^{A_{m,\lambda}^{\pm}}\,,
\ee
where, for $2 \leq m \leq 9$, $A^{\pm}_{m,\lambda}$ is defined as
\bel{Amdef}
A_{m,\lambda}^{\pm} = \frac{m\lambda^{\pm}+1-\lambda^{\pm}}{4m-3}\,.
\ee
In effect, $\lambda^{\pm}$ is a fitting parameter for the maxima in time.  An explanation why such a 
relation should hold can be found in \cite{gibbon_15}. The range of values of $\lambda^{\pm}$ have 
been determined numerically (see Fig.~\ref{Fig:AllRuns} and Table~\ref{table1}). By inserting  (\ref{numdep1}) 
into (\ref{nlt2}) it is found that
\beq{nlt3}
L^{-3}\varpi_{0}^{-3}
\I\left|\bom^{\pm}\cdot\left(\bom^{\mp}\cdot\nabla\bz^{\pm}\right)\right|\,dV &\leq &
c_{4,m} \Dopm^{\frac{\chi_{m}^{\pm}(2m-1) + m(2m-3)}{4m(m-1)}}\Domp^{\frac{\chi_{m}^{\mp} + m(2m-3)}{4m(m-1)}}.
\eeq
Next, the second nonlinear term in (\ref{D1pmA}) is considered where (\ref{nltex1}) is used. From this, it is found 
that the estimate for the right-hand side of (\ref{nlt3}) is the same as in (\ref{nlt1}), apart from the constant $c_{5,m}$\,:
\beq{nlt5}
\sum_{i=1}^{3}\I \left| \bom^{\pm}\cdot\left(\partial_{i}\bz^{\pm}\times\partial_{i}\bz^{\mp}\right)\right|\,dV 
&\leq& c_{5,m}\left(\I|\bom^{\pm}|^{2}\right)^{\frac{2m-3}{4(m-1)}}
\left(\I |\bom^{\pm}|^{2m}\right)^{\frac{1}{4m(m-1)}}\\
&\times& \left(\I |\bom^{\mp}|^{2}\right)^{\frac{2m-3}{4(m-1)}}\left(\I |\bom^{\mp}|^{2m}\right)^{\frac{1}{4m(m-1)}}
\left(\I |\bom^{\pm}|^{2m}\right)^{1/2m}\non\,.
\eeq
Converting this into the $D_{m}^{\pm}$ gives the same right-hand side as in (\ref{nlt3}) but with a constant $c_{2,m}$. Taking 
all these terms together, (\ref{D1pmA}) becomes 
\bel{D1pmB}
\shalf \varpi_{0}^{-1}\dot{D}_{1}^{\pm} \leq - L^{-1}\varpi_{0}^{-2}\I |\nabla\bom^{\pm}|^{2}dV +
c_{6,m} \Dopm^{\frac{\chi_{m}^{\pm}(2m-1) + m(2m-3)}{4m(m-1)}}\Domp^{\frac{\chi_{m}^{\mp} + m(2m-3)}{4m(m-1)}} 
+ Gr_{\pm}D_{1}^{\pm1/2}\,.
\ee
To handle the coupled nature of the $\pm$-variables we define
\bel{X1def1}
X = D_{1}^{\pm} + D_{1}^{\mp}\qquad\mbox{and}\qquad E_{0} = \max_{t}\left(E^{+},E^{-}\right)
\ee
and the two bounded dimensionless energies are defined by $E^{\pm} = \nu^{-2}L^{-1}\I |\bz^{\pm}|^{2}\,dV$. By adding 
the $\pm$-equations and using the depletion formulas (\ref{numdep1}) and (\ref{Amdef}), a differential inequality is found 
for $X(t)$ 
\bel{X1}
\shalf \varpi_{0}^{-1}\dot{X} \leq - \frac{X^{2}}{2E_{0}} + 
c_{6,m} X^{1+ \shalf\lambda^{\pm} - (\lambda^{\pm} - \lambda^{\mp})/4m} + 2\max\left(Gr_{+},\,Gr_{-}\right)X^{1/2}\,.
\ee
Note that when $\lambda^{\pm} = \lambda^{\mp} = \lambda$, as in the Navier-Stokes case, then the exponent of the nonlinear 
term reduces to $1 + \shalf\lambda$, as it should. 
\end{widetext}
Without the use of the numerically observed depletion in (\ref{numdep1}), standard methods in analysis leads to a term $\propto X^{3}$ 
in Eq.(\ref{X1}) (see Ref.~\cite{gibbon_14} for the NS case), which does not lead to a control over the solutions at large times.  However, 
provided $\lambda^{\pm}$ and $\lambda_{\mp}$ satisfy
\bel{ball1}
1+ \shalf\lambda^{\pm} - \frac{(\lambda^{\pm} - \lambda^{\mp})}{4m} < 2 ,
\ee
an `absorbing ball' for $X$ exists because $E_{0}$ is bounded above. This ball has finite radius (depending on the upper bound 
on $E_{0}$) into which solutions are drawn if initial conditions are set outside the ball, and which cannot escape if initial conditions are set 
inside. Expression (\ref{ball1}) can be rewritten as  
\bel{ball2}
\lambda^{\pm} < 2 + \epsilon_{m}^{\pm},
\ee 
where $\epsilon_{m}^{\pm} =  (\lambda^{\pm} - \lambda^{\mp})/2m$, which is a small number. Subject to the constraints on 
$\lambda^{\pm}$ in (\ref{ball2}), the ball is such that the $\bz^{\pm}$ are $L^{2}$-bounded, and thus so are $\bu$ and $\bdb$. Additionally, 
the control of $X$ that (\ref{X1}) affords (an $H_{1}$-bound) is also enough to prove its compactness.  This ball is thus the global attractor 
which governs the long-time dynamics of the PDEs. 

\begin{figure}[!ht]
\includegraphics[width=0.95\columnwidth]{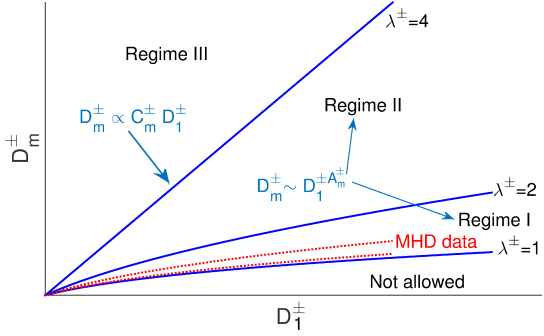}
\caption{(Color online)  Schematic plots of $D_m^\pm$ versus $D_1^\pm$ showing the three 
regimes in 3D MHD (see text). The additive term $\epsilon^{\pm}$ in (\ref{ref1}) has been omitted 
because it is small and can take either sign. The values of $\lambda^{\pm}$ in Table \ref{table1} 
lie only just above the lower bound $\lambda^{\pm} = 1$. Solutions are regular in regime I but 
not in regime II. To start in regime III requires unphysical initial data.}\label{Fig:Cartoon}
\end{figure}

\begin{figure*}
\includegraphics[width=0.99\textwidth]{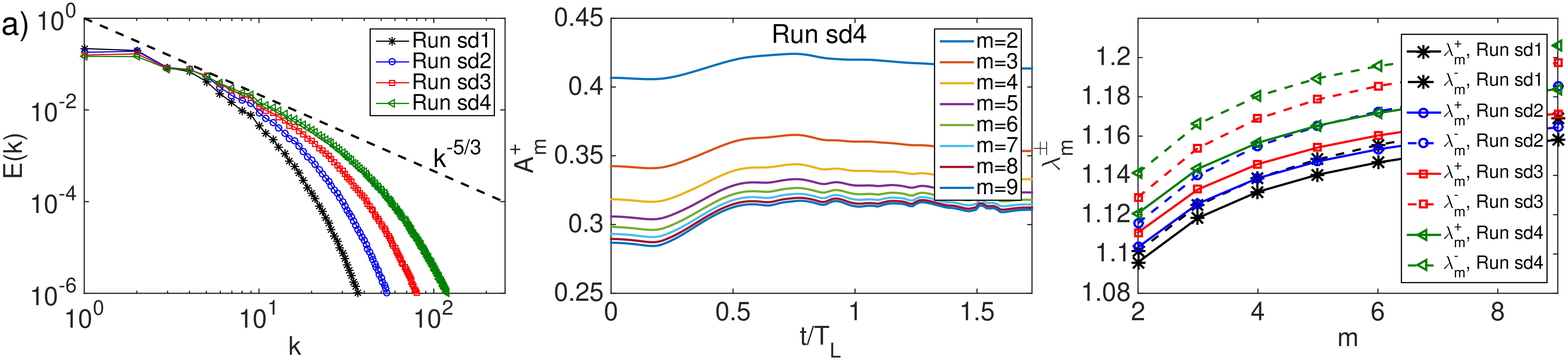}
\includegraphics[width=0.99\textwidth]{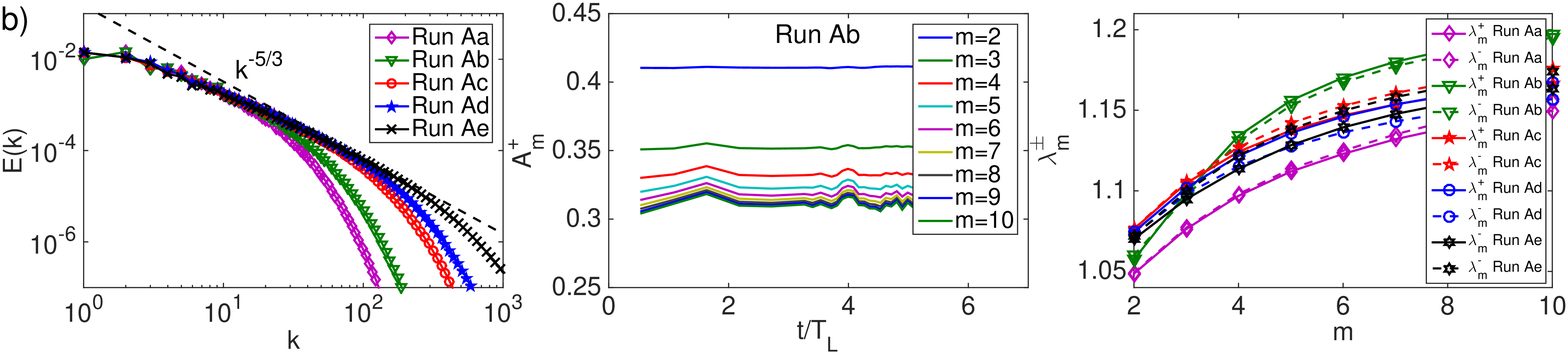}
\includegraphics[width=0.99\textwidth]{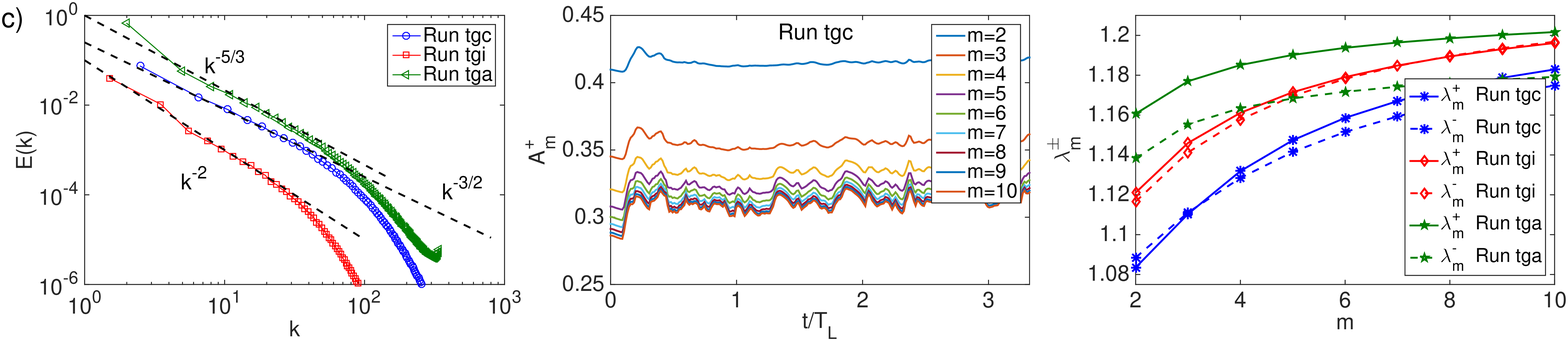}
\includegraphics[width=0.99\textwidth]{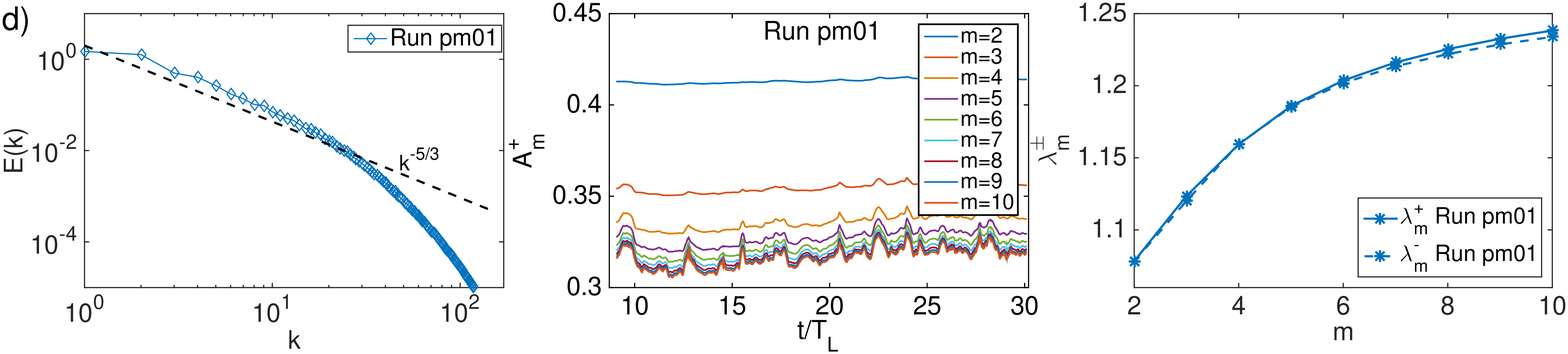}
\caption{(Color online) Representative results from our DNSs:  total energy spectra (first column), temporal evolution 
of $A_m^+$ (second column) and values of $\lambda_m^\pm$ (third column). The rows correspond to different runs\,: 
a) decaying 3D MHD turbulence\,; b) forced, statistically steady 3D MHD turbulence\,; c) forced, statistically steady 3D MHD 
turbulence with imposed Taylor-Green symmetries and d) forced, statistically steady 3D MHD turbulence with $P_M=0.1$. 
For parameters see Table \ref{table1}.}
\label{Fig:AllRuns}
\end{figure*}

The \textit{natural}, 3D-MHD analogs of Fig. 1 in Ref.~\cite{gibbon_14} are the schematic plots of $D_m^+$ versus $D_1^+$ and 
$D_m^-$ versus $D_1^-$ in Fig.~\ref{Fig:Cartoon}, which show three regimes. For regular solutions, we must have 
\bel{ref1}
1 \leq \lambda^{\pm} \leq 2+\epsilon^{\pm}\,,\qquad\mbox{(regime I)}\,.
\ee
The $\epsilon^{\pm}$-term has been left off the figure as it is small and can take either sign. When 
\bel{reg2}
2+\epsilon^{\pm} \leq \lambda^{\pm} < 4\,,\qquad\mbox{(regime II)}
\ee
there is depletion, but not enough to control solutions; and, finally, when 
\bel{reg3}
\lambda_{\pm} \geq 4\,,\qquad\mbox{(regime III)}
\ee
then $D_m^{\pm} \geq c_{m} D_1^{\pm}$. However, any initial data set in this region would be pathological as it would 
have to be prepared as a very large spike in $\bom^{\pm}$ in which the $L^{\infty}$-norm is much larger than the $L^{2}$-norm. 
In the NS-case, it can be shown that solutions are regular in regime III, but no more than algebraically increasing  because of the 
forcing \cite{gibbon_15}. 
\hspace{2mm}
\textit{Our DNS data indicate that regime I ($1 \leq \lambda^\pm \leq 2 + \epsilon^{\pm}$) is obtained in 3D MHD 
for all the solutions we have studied.}

In Fig.~\ref{Fig:AllRuns} representative results from four of our DNSs are  given. The  first column of 
Fig.~\ref{Fig:AllRuns} contains log-log (base 10) plots of the energy spectra $E(k)$ 
versus $k$. Most of these energy spectra show power-law forms in the inertial range with an exponent that is 
consistent with the K41 value $-5/3$. However, this exponent is consistent with the IK value $-3/2$ for run-tgc\,; 
and it is $-2$ for run-tgi. We find that these exponents can depend on the values of $\sigma_{C}$ and $\sigma_{M}$, 
which are given in Table \ref{table1}.  

The second column of Fig.~\ref{Fig:AllRuns} has plots of $A_m^+(t)$ versus $t$, from which $A_{m,\lambda}^+$ is 
determined (plots for $A_m^-(t)$ versus $t$ are similar), which follow from $D_m^+(t)$. 
The region where the data lie do not follow exactly the contour boundary curves of Fig.~\ref{Fig:Cartoon}. The $\lambda^{\pm}$ 
have been determined as in Ref.~\cite{gibbon_14} for the 3D-NS equations.  $\lambda_{m}^{\pm}$ are defined to be those values 
that have been computed from Eq.~(\ref{Amdef}) for $A_{m,\lambda}^{\pm}$.  A check has shown that our data are reliable 
up to $m=10$\,; note the ordering of the $A_{m,\lambda}^{\pm}$ is the same for all our runs. 

In the third column of Fig.~\ref{Fig:AllRuns}, plots of $\lambda_{m}^{\pm}$ versus $m$ are given,  in the range $2 \leq m 
\leq 9$, for which good-quality numerical data have been obtained. From these plots $\lambda^{\pm}$ has been found 
from the minimum over $m$ of $\lambda_{m}^{\pm}$. In general, $1\le \lambda_\pm \le 4$; however, in all our DNSs, 
$1\le \lambda_\pm \le 2$, i.e., our solutions lie in regime I. 
\begin{figure}[!ht]
\includegraphics[width=0.4\textwidth]{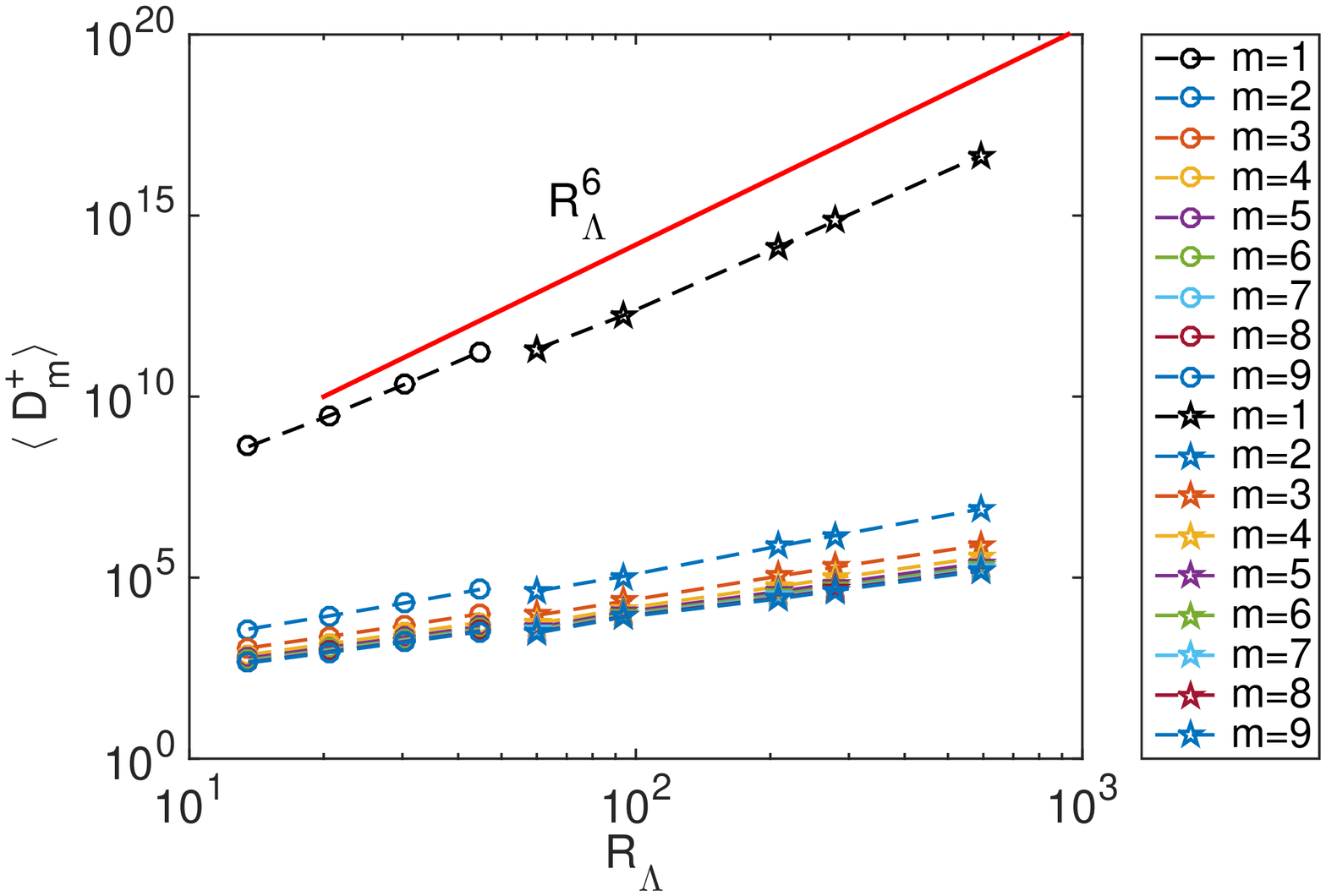}
\caption{(Color online) Plots versus $R_\Lambda$  of $\left< D_{m}^{+}\right>_{st.av.}$, 
for the runs Aa-Ae  (pentagrams) and its decaying-MHD analog, for the runs 
sd1-sd4 (circles), where we use the value of $D_m^+$ at the time at which the 
energy dissipation rate $\epsilon$ reaches its first maximum~\cite{sahoo_11}. 
See Table \ref{table1} for additional information about these runs.}\label{Fig:DpVsRe}
\end{figure}

\vspace{5mm}

\section{Spectra\label{sec:spectra}}


\subsection{How to estimate the spectrum for the 3D MHD-Els\"asser system\label{EST:Spec}}

The method of Doering and Gibbon \cite{doering_02b} is now followed which explains how to estimate 
average length scales and a corresponding spectrum based on ideas in \cite{frisch_95}. It is necessary 
to define a set of time-averaged inverse length scales\footnote{For technical reasons \cite{doering_02b}, 
an additive term should be included in both the denominator and numerator in (\ref{ls1}) to take account 
of the forcing term in (\ref{D1pmB}), but as this makes a negligible contribution it will be dropped.} 
\begin{widetext}
\beq{ls1}
\left<L^{2}\kappa_{2,1}^{\pm2}\right>_{T} &=& \left<\frac{L^{2}\|\nabla\bom^{\pm}\|_{2}^{2}dV}{\|\bom^{\pm}\|_{2}^{2}}\right>_{T}
=\left<\frac{L^{-1}\varpi_{0}^{-2}\I |\nabla\bom^{\pm}|^{2}dV}{D_{1}^{\pm}}\right>_{T}\,,
\eeq
where the labeling of the subscripts is based on the number of derivatives on $\bz^{\pm}$. Dividing (\ref{D1pmB}) by 
$D_{1}^{\pm}$ and time averaging, we find\footnote{As a check we note that if $\chi_{m,\lambda}^{+}=\chi_{m,\lambda}^{-} 
= m\lambda + 1-\lambda$, as it does in the pure $3D$ Navier-Stokes case \cite{gibbon_14}, then the sum of the exponents 
on the right hand side of (\ref{ls2}) is $\shalf \lambda$, as it should.}
\beq{ls2}
\left<L^{2}\kappa_{2,1}^{\pm2}\right>_{T} &\leq& 
c_{6,m}\left<\Dopm^{\frac{\left(\chi_{m}^{\pm}-m\right)(2m-1)}{4m(m-1)}}
\Domp^{\frac{\chi_{m}^{\mp} + m(2m-3)}{4m(m-1)}}\right>_{T}\nonumber\\
&\leq& c_{6,m}\left<D_{1}^{\pm}\right>_{T}^{\frac{\left(\chi_{m}^{\pm}-m\right)(2m-1)}{4m(m-1)}}
\left<D_{1}^{\pm}\right>_{T}^{\frac{\chi_{m}^{\mp} + m(2m-3)}{4m(m-1)}}.
\eeq
\end{widetext}
Secondly, it is necessary to estimate $\left<D_{1}^{\pm}\right>_{T}$ by using the energy inequality version of (\ref{zfielda}) 
\bel{D1estA}
\shalf\frac{d~}{dt} \|\bz^{\pm}\|_{2}^{2} \leq - \nu \|\bom^{\pm}\|_{2}^{2} + \|\bz^{\pm}\|_{2}\|\bdf^{\pm}\|_{2}\,.
\ee
By time averaging, converting into a dimensionless form, and using (\ref{GrRe}), it is found that
\beq{D1estB}
\left<D_{1}^{\pm}\right>_{T} \leq Gr_{\pm}Re_{\pm} \leq c\,Re_{\pm}^{2}\left(Re_{\mp} + 1\right)\,.
\eeq
Moreover, by introducing the definitions, the first of which is the Els\"asser analog of the Taylor micro-scale, 
\bel{k10def1}
\kappa_{1,0}^{\pm2} =  \frac{\|\bom^{\pm}\|_{2}^{2}}{\|\bz^{\pm}\|_{2}^{2}}\,,\qquad
\kappa_{2,0}^{\pm4} = \frac{\|\nabla\bom^{\pm}\|_{2}^{2}}{\|\bz^{\pm}\|_{2}^{2}}\,,
\ee
and adapting ideas in \cite{doering_02b}, (\ref{D1estA}) gives 
\bel{ls4}
\left<L^{2}\kappa_{1,0}^{\pm2}\right>_{T} \leq Re_{\pm}\,.
\ee 
Then it is easily shown that
\bel{ls5A}
\left<L\kappa_{2,0}^{\pm}\right>_{T} \leq \left<L^{2}\kappa_{2,1}^{\pm2}\right>_{T}^{1/4}
\left<L^{2}\kappa_{1,0}^{\pm2}\right>_{T}^{1/4}\,,
\ee
and so from (\ref{ls1}), (\ref{ls4}) and (\ref{D1estB}), in which only the dominant term has been kept, it is found that
\beq{ls5B}
\left<L\kappa_{2,0}^{\pm}\right>_{T} 
\leq c_{6,m}Re_{\pm}^{\sigma_{m}^{\pm}}\left(Re_{\mp}+1\right)^{\rho_{m}^{\mp}}\,,
\eeq
where, with $\chi_{m}^{\pm} = (m-1)\lambda^{\pm} +1$, 
\beq{sigdef1}
\sigma_{m}^{\pm} &=& \frac{\lambda^{\pm} + 1}{4} + \left(\frac{\lambda^{\mp} - \lambda^{\pm}}{8m}\right)\,,\\
\rho_{m}^{\mp} &=& \frac{(2m-1)\lambda^{\mp} + \lambda^{\pm}}{16m}\,.
\eeq
Thus, (\ref{sigdef1}) can be written as 
\bel{sigdef2}
\left<L\kappa_{2,0}^{\pm}\right>_{T} \leq c\,Re_{\pm}^{\frac{\lambda^{\pm} + 1}{4}}Re_{\mp}^{\frac{\lambda^{\mp}}{8}}\,,
\ee
where the factor of unity has been ignored in the large $Re^{\pm}$-limit and the limit of large $m$ has been taken. 
Then the problem is whether it is possible to consider $Re_+$ and $Re_-$ as independent variables or not. 
The simplest way is to note that
\beq{sigdef3}
\|\bz^\pm\|^{2} 
&\leq& \|\bv\|^{2}_{2}  + 2 \|\bv\|_{2}\|\bdb\|_{2} + \|\bdb\|^{2}_{2} \non\\
&\leq& 2\left(\|\bv\|^{2}_{2}  + \|\bdb\|^{2}_{2}\right) = 4 E_{\rm tot}.
\eeq
Defining a \textit{global} Reynolds number as 
\bel{globRe}
Re=L \sqrt{2 E_{\rm tot}}/\nu\,,
\ee
(\ref{sigdef2}) gives
\bel{sigdef4}
\left< L \kappa^\pm_{2,0} \right>_{T} \leq c\, Re^{\frac{\lambda^\pm+1}{4}+\frac{\lambda^\mp}{8}}\,.
\ee
Now the implications of these results are examined for energy spectra derived in (\ref{sigdef4}). By assuming 
isotropy and the power-law Ans\"atze
\bel{app3}
\mathcal{E}^{\pm}(k) = \left\{ 
\begin{array}{cr}
A\,k^{-q^{\pm}}, & \ \ \ L^{-1}\leq k \leq k_{c}^{\pm}\,,\\
0, &       k > k_{c}^{\pm}\,,
\end{array}\right.
\ee
the identification (see Appendix B and \cite{frisch_95}) 
\bel{Eq:kappa2Sim}
\left<L\kappa_{2,0}^{\pm}\right>_{T}\sim (L\,k_{c}^{\pm})^{1-\frac{q^{\pm}-1}{4}}
\sim Re^\frac{5-q}{4(3-q)}\,,
\ee
allows us to find an inequality relation between $q^{\pm}$ and $\lambda_{\pm}$. 
\bel{qineq}
q^\pm\ge 3-\frac{4}{2\lambda^\pm+\lambda^\mp}\ge\frac{5}{3},
\ee
which excludes the IK exponent $3/2$, at least in the absence of intermittency. 
However, full isotropy has been assumed at all scales and the limit $Re\to\infty$ 
when comparing \eqref{sigdef4} and \eqref{Eq:kappa2Sim}. Furthermore, 
\eqref{Eq:kappa2Sim} implicitly assumes $E_+ = E_-$. In general, this relation 
is modified leading to a set of inequalities for the $q^\pm$ that do not exclude IK 
(the last two equations in Appendix B). In fact, IK is excluded \textit{only} if correlation 
is neglected. This is consistent with the fact that run-tgc (see Fig. 2 and Table I) 
produces an IK scaling because it has a non-zero cross-correlation. It is known 
that in the presence of cross-correlations between the velocity and magnetic 
field ($H_C\not= 0$), 
different indices arise for the energy spectra of the two Els\"asser fields, 
$\bz^\pm$ (see, e.g., \cite{pouquet_88} and references therein), a result that 
persists in the case of weak MHD turbulence, as shown through wave-turbulence 
developments \cite{galtier_00}. The mathematical analysis as well as the numerical 
simulations presented in this paper, all put on firm ground that indeed $H_C$ plays 
a crucial role in determining the distribution of energy among scales.

In Fig.~\ref{Fig:DpVsRe} plots of $\left< D_{m}^{+}\right>_{st.av.}$ versus $R_\Lambda$ have 
been displayed, where the angular brackets now indicate the average over the statistically 
stationary turbulent state\,; we present data (pentagrams in Fig.~\ref{Fig:DpVsRe}) from the 
runs Aa-Ae (Table \ref{table1}). Circles indicate decaying-MHD data points (runs sd1-sd4); 
here we use the value of $D_m^+$ at the time at which the energy-dissipation rate $\epsilon$ 
reaches its first maximum~\cite{sahoo_11}. The order-$m$ moments of gradients, or 
\textit{gradmoments}, of hydrodynamic fields have been used to investigate Nelkin 
scaling~\cite{nelkin90,schumacher07,chakraborty12}, i.e., the power-law dependence
of the  gradmoments on the Reynolds number $Re$ in the case of fluid turbulence; the 
Nelkin-scaling exponents $\xi_m$ can be related to the structure-function exponents 
$\zeta_m$~\cite{nelkin90,schumacher07,chakraborty12}.  Our vorticity moments are upper 
bounds for gradmoments of the Els\"asser variables (see (\ref{nltex1})). If these bounds are 
saturated, then the exponents, which can be extracted from the plots of Fig.~\ref{Fig:DpVsRe},
should be related to the Nelkin exponents for 3D MHD turbulence. A more detailed exposition 
of such scaling in 3D MHD has been deferred to another study. 

In liquid metals, as well as in the solar photosphere, $P_M$ is very small. It can also be very large as, e.g., 
in the interstellar medium. Our mathematical analysis is not valid if $P_M \neq 1$ because $\nu_-$ can 
become negative for $P_M \leq 1$. However, our DNS results in Fig.~\ref{Fig:DpVsRe} show that plots for 
$P_M = 0.1$ (bottom row) are similar to their counterparts for $P_M=1$ (top three rows). Furthermore, 
at least for a fixed value of $\nu$, Table \ref{table1} shows that the values of $\lambda_m^\pm$ are 
comparable to their $P_M = 1$ counterparts\,; these values of  $\lambda^\pm$ decrease marginally 
as $P_M$ is increased.



\section{Conclusions}\label{Con}

Our work, which builds upon the studies of Refs.~\cite{gibbon_12,donzis_13,gibbon_14} for fluid turbulence, 
provides insights into the depletion of nonlinearity in 3D MHD turbulence and its intermittency. In 
particular, we have introduced the scaled moments $D_m^{\pm}$, and then obtained inequalities containing 
$D_m^{\pm}$ and $D_1^{\pm}$\,; these inequalities specify three possible regimes. In essence, it has been 
found that 3D MHD turbulence is similar to its fluid-turbulence counterpart insofar as all solutions that have 
been investigated have remained in only one regime (regime I), which displays depleted nonlinearity 
(Fig.~\ref{Fig:Cartoon}).  Moreover, under the assumption of isotropy our results lead to the inequality~(\ref{qineq}) 
for the spectral exponents $q^{\pm}$. In fact, the inequality (\ref{nltex1}) can relate $D_m^{\pm}$ and the order-$m$ 
moments of gradients of the magnetohydrodynamic fields\,; such moments can then be used, along with a suitable 
generalization of Nelkin scaling~\cite{nelkin90,schumacher07,chakraborty12} for 3D MHD turbulence, to
relate slopes of plots like those in Fig.~\ref{Fig:DpVsRe} to the multiscaling exponents of Els\"asser-field 
structure functions. We conclude that 3D MHD appears to have more nonlinear depletion than fluid 
turbulence, because the values of $\lambda^\pm$ are lower than those for their fluid-turbulence 
counterparts\,; this can be attributed to Alfv\'en waves weakening the nonlinear eddies.

\begin{acknowledgments} 
We thank H. Homann and R. Grauer for providing the  data of runs Ad and Ae. Data from runs Ad and Ae 
of H. Homann and R. Grauer were produced on the IBM BlueGene/P computer JUGENE at FZ J\"ulich 
made available through the ``XXL project of HBO28''. We also acknowledge U. Frisch and D. Vincenzi for 
useful discussions. The computations for runs sd1-sd4 were performed on Janus (UC-Boulder). Runs Aa-Ac 
and tg were performed at M\'esocentre SIGAMM hosted at the Observatoire de la C\^{o}te d'Azur and 
CICADA hosted by the University of Nice-Sophia. Computer time was also provided by GENCI on the 
IDRIS/CINES/TGCC clusters. JG and AP thank F\'ed\'eration Doeblin for support. RP thanks the Department 
of Science and Technology (India) for support and SERC (IISc) for computational resources.  AG is grateful 
for support through a  grant from the European Research Council (ERC) under the European Community's 
Seventh Framework Programme (FP7/2007-2013)/ERC Grant Agreement No. 297004.  GS acknowledges 
support from the ERC Advanced Grant ``NewTURB'', No. 339032. GK thanks the Indo-French Centre for 
Applied Mathematics (IFCAM) for supporting a visit during which parts of this paper were written. JS is 
supported by the National Science Foundation (NSF) Graduate Research Fellowship Program (GRFP) 
under Grant No. DGE 1144083. AP and JDG acknowledge, with thanks, IPAM UCLA where this collaboration 
began in the Autumn of 2014 on the program ``Mathematics of Turbulence''.
\end{acknowledgments}

\appendix
\begin{widetext}
\section{Vorticity and current equations}\label{app1}

The following four identities have been used for vectors $\bdD$ and $\bdQ$ (not necessarily divergence-free)\,: 
\begin{eqnarray}\label{eq:identi} 
\nabla \times [\nabla \times \bdD] &=& \nabla [ \nabla \cdot \bdD] - \nabla^{2} \bdD \,\non\\
\nabla  [\bdD \cdot \bdQ] &=& \bdD \times [\nabla \times \bdQ] + \bdQ \times  [\nabla \times \bdD]\non\\ 
&+& \bdD \cdot \nabla \bdQ + \bdQ \cdot \nabla \bdD\,\non\\
 \nabla \cdot [\bdD \times \bdQ] &=&\  \bdQ \cdot \nabla \times \bdD - \bdD \cdot  \nabla \times \bdQ\,\non\\
\nabla \times [\bdD \times \bdQ] &=& \bdQ \cdot \nabla \bdD - \bdD \cdot \nabla \bdQ\,. 
\end{eqnarray}
The equation for the vorticity is derived straightforwardly\,:
\bel{eq:ome}
(\partial_{t} + \bu \cdot \nabla) \bom = \bom\cdot \nabla \bu + \bdb \cdot  \nabla \bj - \bj \cdot  \nabla \bdb\,.
\ee
By using $\bdD = \bu \times \bdb$ in the above identities (with $\nabla \cdot \bdD \neq 0$),  we  obtain the 
equation for the current
\bel{curr1}
\partial_{t} \bj = \nabla\left[ \nabla \cdot [\bu \times \bdb] - \nabla^2[ \bu \times \bdb ]\right]\,,
\ee
which, upon expansion, leads to\,:
\beq{curr2}
 \nabla [\nabla \cdot [\bu \times \bdb] ] &=& \bom \cdot \nabla \bdb + \bdb \cdot \nabla \bom - \bdb \times \nabla^{2}\bu + \bom \times \bj
                                                                           -  \bu \cdot \nabla \bj - \bj \cdot \nabla \bu + \bu \times \nabla^{2} \bdb - \bj   \times \bom\\
- \nabla^{2}[ \bu \times \bdb ]  &=& - \nabla^{2} \bu \times \bdb   -  \bu \times \nabla^{2}\bdb
-2 \Sigma_{i}\partial_{i}\bu \times \partial_{i}\bdb \,.
 \eeq
 So  the equation for the current  (note the cancellations in the $\nabla^2$ terms) is:
 \bel{curr3}
(\partial_{t} + \bu \cdot \nabla)\bj -  \bom\cdot \nabla \bdb - \bdb \cdot \nabla \bom + \bj \cdot \nabla \bu  
- 2 \bom \times \bj  =   -2 \Sigma_{i}\partial_{i}\bu \times \partial_{i}\bdb  \ ,
 \ee
 an expression already written in \cite{sanminiato}. 
 For the curl of the Els\"asser field $\bom^{+}=\bom +\bj$, (with $\pm$ symmetry for  $\partial_t \bom^{-}$), 
(\ref{curr3}) and (\ref{eq:ome}) reduce to (\ref{omz})
\bel{eq:ompm2} 
(\partial_{t} + \bz^{-} \cdot \nabla) \bom^{+}  =   \bom^{-} \cdot \nabla\bz^{+} + \bom^{-} \times \bom^{+}  + 
\Sigma_{i} \partial_{i}\bz^{+} \times \partial_{i}\bz^{-} \,.
\ee
\end{widetext}
Note that the geometry term $2 \bom \times \bj$ in the equation for the current density does not appear in the vorticity equation; 
also, it is weak for almost-aligned current and vorticity (or $\bom^\pm$) \cite{servidio_09, stawarz_12}. The equations (\ref{eq:ompm2}) 
for the temporal evolution of $\bom^{\pm}$ follow immediately from the above. Note also that $\bom^{-} \times \bom^{+}  = 2 \bom 
\times \bj $ does not affect the point-wise production of $\bom^{\pm}$, whereas the second term can create current density\,; 
here the labels $i = 1,\,2$, and $3$ refer respectively to the $x,\,y$, and $z$ derivatives. Finally note that, for a flow evolving towards 
strong local correlations between the velocity and magnetic field ($\bz^{+}=0$ or $\bz^{-}=0$), this extra term is weak.

\section{Phenomenological argument for fluids and MHD} \label{P:F}

In the fluid case, the total energy and dissipation can be written in terms of the energy spectrum with spectral index $q$ as
\bel{C1}
U^2 = \int_{k_0}^{k_c} A k^{-q}\,,\qquad \epsilon = \nu \int_{k_0}^{k_c} A k^{2-q}\,,
\ee
with the dimension of $A$ as $[A] = [\epsilon^a ] [L^b]$. One finds straightforwardly $a=2/3, \ b= [5-3q]/3$, so $b=0$ for $q=5/3$, 
as expected. This leads to a cut-off wavenumber $k_{c}/k_{0} = [\epsilon \nu^{-3}] ^{1/[3(3-q)]} \ k_0^{-4/[3(3-q)]}$, or in terms 
of the Reynolds number,
\beq{C2}
Re &=&UL/\nu =\epsilon^{1/3}L^{4/3}\nu^{-1}\,,\qquad L k_c = Re^x, \non\\
x &=& [3-q]^{-1}\,,
\eeq
with $k_0 =2\pi/L$. 

In MHD, one can follow the weak-turbulence IK prescription (remaining in the isotropic framework for simplicity). Then, 
$A= [\epsilon B_0]^c L^d$, where $B_0$ is a large-scale strong (quasi)-uniform magnetic field; so $c=1/2,~d=[3-2q]/2$.
This leads to
\bel{C3}
k_{c}/k_{0} = [\epsilon B_0^{-1} \nu^{-2}] ^{1/[2(3-q)]} \ k_0^{-3/[2(3-q)]}\,;
\ee
or, in terms of the Reynolds number,
\bel{C4}
Re=UL/\nu= [\epsilon B_0 L^5]^{1/4} \nu^{-1}\,,
\ee
with
\beq{C5}
L k_c &=& r^x\,Re^x,\qquad x= [3-q]^{-1}\,,\non\\
 r &=&\frac{U}{B_0} << 1   \ ,    
\eeq
which is a hypothesis that is compatible with the wave-turbulence assumption. Thus, with the introduction of the factor $r$, the scale 
dependence of the cut-off wavenumber with Reynolds number is the same for fluids and MHD.

This phenomenological argument can be reproduced in the more general case when the velocity and magnetic fields are correlated, i.e.,  
with $E_{+}\not= E_{-}$. This results in a condition between the indices $q_\pm$ of the $E_\pm$ spectra, namely $q_{+}+q_{-} =3$ within 
the IK framework with, as before for the uncorrelated flows, $q_{+}=q_{-} =3/2$ (see \cite{sanminiato} for an introduction). Two-point closure 
computations and two-dimensional numerical simulations find $q_+\not= q_-$ at high correlations, but the three-dimensional case 
remains open. As in the preceding case of uncorrected MHD, we make the assumption that the $\pm$ integral scales are both comparable 
to the box size $L$.

After some algebra along the same lines as before, one finds that the dissipative wave numbers for the $E_\pm(k)$ spectra are equal both to 
$k_{+}=k_{-}=k_{c}=\epsilon/[\nu^2 B_0]^{1/3}$, as in the zero-correlation case, or\,:
\bel{C6} 
L k_\nu = [z_0^+/B_0]^{1/3} [z_0^-/B_0]^{1/3} Re_+^{1/3} Re_-^{1/3} \,, 
\ee 
with $Re_\pm=z_0^\pm L/\nu$, and we have assumed that the magnetic Prandtl number is equal to unity so that $\nu=\eta=\nu_+=\nu_-$. Writing 
\beq{C7} 
E_\pm(k)&=&A_\pm(\epsilon, B_{0},\,L)k^{-q_\pm}\,,\non\\
A_\pm(\epsilon, B_{0},\,L) &=& \epsilon^{a_\pm} B_0^{b_\pm}L^{c_\pm} \,, 
\eeq
it is readily found that, under the assumption that 
\bel{C8a}
b_{+} + b_{-} =-3(a_++a_-)+4=1
\ee
(so that $E_+(k)E_-(k)\sim [\epsilon B_0]k^{-3}$), and that 
\bel{C8b}
c_{+} + c_{-} = -3+(3-q_+)+(3-q_-)\,,
\ee
then
\bel{C9}
L k_c \sim \frac{\epsilon}{[\nu^2 B_0]}^{\frac{1}{(3-q_+)+(3-q_-)}} \, L^{\frac{3}{(3-q_+)+(3-q_-)}} \,, 
\ee
or, in terms of the Reynolds numbers $Re_{\pm}$,
\beq{C10} 
L k_{c} &\sim& [[z_0^+/B_0][z_0^-/B_0]]^{\frac{1}{(3-q_+)+(3-q_-)}}\non\\
&\times&[Re_+ Re_- ]^{\frac{1}{(3-q_+)+(3-q_-)}} \,.
\eeq
Finally, this phenomenological relation can be used as in the non-correlated case to establish the upper bounds 
of spectral indices. In this case one finds that
\bel{C11}
\frac{\left[1 - \squart(q_{+}-1)\right]}{(3-q_+) + (3-q_-)} \leq \min \left\{\frac{\lambda_{+}+1}{4}\,;\frac{\lambda_-}{8} \right\}\,,
\ee
and similarly
\bel{C12}
\frac{\left[1-\squart(q_{-}-1)\right]}{(3-q_+) + (3-q_{-})} \leq \min \left\{ \frac{\lambda_-+1}{4}\,;\frac{\lambda_+}{8} \right\}\,,
\ee
with $1< q_\pm < 3$ and $1+\lambda_\pm/2 < 2$ using equation (\ref{ball2}). These results are to be contrasted with the 
uncorrelated case obtained in the Kolmogorov framework. It is possible to show that in the correlated case the IK spectrum 
$q_\pm=3/2$ cannot be excluded.


\section{The Doering-Foias $Gr^{\pm}-Re^{\pm}$ relation for MHD}\label{app2}

Following Doering and Foias \cite{doering_02} the forcing function $\bdf^{\pm}(\bx)$ is split  into its magnitude $F^{\pm}$ 
and its ``shape'' $\bphi^{\pm}$ such that
\bel{p1}
\bdf^{\pm}(\bx) = F^{\pm}\bphi^{\pm}(\ell^{-1}\bx),
\ee
where $\ell$ is the longest length scale in the force and is taken to be $\ell = L$ for convenience in the rest of the paper. 
On the unit torus $\mathbb{I}_{d}$ in $d$-dimensions, $\bphi$ is a mean-zero, divergence-free vector field with the 
chosen normalization property 
\bel{p1a}
\int_{\mathbb{I}_{d}} \left|\nabla^{-1}_{y}\bphi^{\pm}\right|^{2}\,d^{d}y = 1\,.
\ee
$L^{2}$-norms of $\bdf^{\pm}$ on $\mathbb{I}^{d}$ are 
\bel{p2}
\|\nabla^{N}\bdf^{\pm}\|_{2}^{2} = C_{N}^{\pm}\ell^{-2N} L^{d} F^{\pm2},
\ee
where the coefficients $C_{N}^{\pm}$, which refer to the shape of the force but not its magnitude, are 
\bel{p3}
C_{M}^{\pm} = \sum_{n} \left| 2\pi n\right|^{2N} |\hat{\bphi}^{\pm}_{n}|^{2}\,.
\ee
Various bounds exist such as (among others)
\bel{p4}
\|\nabla\Delta^{-M}\bdf^{\pm}\|_{\infty} = D_{M}^{\pm}F \ell^{2M-1}\,.
\ee
The energy dissipation rate $\epsilon$ is  
\bel{p5}
\epsilon^{\pm} = \left<\nu L^{-d}\I |\nabla\bz^{\pm}|^{2}\,dV\right> = \nu L^{-d}\left<H_{1}^{\pm}\right>\,.
\ee
In terms of $F^{\pm}$, the Grashof number in (\ref{Grdef1}) becomes ($\ell = L$)
\bel{Grdef2}
Gr_{\pm} = F^{\pm}\ell^{3}/\nu^{2} \ .
\ee
Following the procedure in \cite{doering_02} (pg 296 equation (2.9)), we multiply (\ref{zfielda}) by 
$(-\Delta^{-M})\bdf^{\pm}$  and integrate to obtain 
\begin{widetext}
\beq{a1}
\frac{d~}{dt}\int_{\mathbb{I}_{d}} \bz^{\pm}\cdot[(-\Delta^{-M})\bdf^{\pm}]\,dV 
&=& \nu\int_{\mathbb{I}_{d}} \Delta\bz^{\pm}\cdot\left[(-\Delta^{-M})\bdf^{\pm}\right] 
- \int_{\mathbb{I}_{d}} \bz^{\mp}\cdot\nabla\bz^{\pm}\cdot\left[(-\Delta^{-M})\bdf^{\pm}\right]\,dV \nonumber\\
&+&\int_{\mathbb{I}_{d}}\bdf^{\pm}\cdot\left[(-\Delta^{-M})\bdf^{\pm}\right]\,dV\,.
\eeq
Now if we integrate all the terms by parts, and take the time average, we get
\beq{a2}
\left<L^{-d}\int_{\mathbb{I}_{d}}\left|\nabla^{-M}\bdf^{\pm}\right|^{2}\,dV\right> &\leq& 
\nu\left<L^{-d}\int_{\mathbb{I}_{d}}\left|\bz^{\pm}\cdot(-\Delta^{-M+1})\bdf^{\pm}\right|\,dV\right>\nonumber\\
&+&\left<L^{-d}\int_{\mathbb{I}_{d}}\left|\bz^{\mp}\cdot[\nabla[(-\Delta^{-M})]\bdf]\cdot\bz^{\pm}\right|\,dV\right> \ .
\eeq
Thus, after a Schwarz inequality, (\ref{a2}) turns into 
\bel{a4}
c_{0}F^{\pm2}\ell^{2M}\leq c_{1}\nu F^{\pm}\ell^{2M-2}U^{\pm} + c_{2}\ell^{2M-1}F^{\pm} U^{\pm}U^{\mp}\,.
\ee
By using (\ref{Grdef2}), in the limit $Gr_{\pm} \to\infty$, (\ref{a4}) becomes 
\bel{a5}
Gr_{\pm} \leq c\,\left(Re_{\pm} + Re_{\pm}Re_{\mp}\right)\,.
\ee
\end{widetext}

\bibliographystyle{apsrev}
\bibliography{Biblio}

 \end{document}